\documentclass{elsart3p}

\usepackage{graphicx}
\usepackage{amsmath}

\begin{document}

\begin{frontmatter}

\title{The two-orbital Hubbard model and the OSMT}

\author[a]{Adolfo Avella}
\author[a]{Ferdinando Mancini}
\author[b]{Satoru Odashima}
\author[a]{Giovanni Scelza\corauthref{Scelza}}
\ead{scelza@sa.infn.it}
\ead[url]{http.//www.sa.infn.it/Homepage.asp?scelza}
\corauth[Scelza]{Corresponding author}
\address[a]{Dipartimento di Fisica ``E.R. Caianiello'' - Unit\`a CNISM
di Salerno \protect\\ Universit\`a degli Studi di Salerno,  I-84081
Baronissi (SA), Italy}
\address[b]{JST Satellite Iwate, 3-35-2 Iiokashinden, Morioka, Iwate 020-0852, Japan}
\begin{abstract}
We analyze the two-orbital Hubbard model by means of the Composite
Operator Method with the aim at studying the phenomenon of orbital
selective Mott transition (OSMT). The model contains an
interorbital interaction $U'$, in addition to the usual
intraorbital one $U$. As warming-up approximation, we use a basis
of two operators only, the Hubbard operators. The analysis of the
density of states at the chemical potential as a function of the
ratio between the bandwidths of the two orbitals shows the clear
signature of an orbital selective Mott transition as expected.
\end{abstract}


\end{frontmatter}
The two-orbital Hubbard model has recently come into the limelight
as a toy model for the study of a phenomenon that seems to
interest a certain number of materials
\cite{Anisimov_02,Medici_05,Ferrero_05}: the orbital selective
Mott transition. In a system composed of two electronic species (a
two orbital system) is possible that, under the influence of
strong electronic correlations, one of the two orbital becomes
insulating, while the other stays metallic as the whole system,
obviously. In this manuscript, we present a preliminary study of
what of this physics the Composite Operator Method
\cite{Mancini_04} is capable to grasp within a simple two-pole
approximation. We consider the following two-orbital Hubbard
model:
\begin{align}
H&=-2d\sum_{\mathbf{i},a}t^{(a)}c^\dag_a(i)c^\alpha_a(i)
-\mu\sum_{\mathbf{i},a}c^\dag_a(i)c_a(i) \nonumber\\
&+U\sum_{\mathbf{i},a}D_a(i) +U'\sum_{\mathbf{i}}n_1(i)n_2(i)
\end{align}
where $c^\dag_{a,\sigma}(i)$ and $c_{a,\sigma}(i)$ are,
respectively, creation and annihilation electron fields with spin
$\sigma(=\uparrow,\downarrow)$ and orbital index $a(=1,2)$,
satisfying anticommutation canonical relations. $\mathbf{i}$ stands
for the lattice vector $\mathbf{R}_i$ and $i=(\mathbf{i},t)$.
$n_{a,\sigma}(i)=c^\dag_{a,\sigma}(i)c_{a,\sigma}(i)$ is the
particle density operator of electrons of spin $\sigma$ and orbital
index $a$. $U$ and $U'$ are the intraorbital and interorbital
Coulomb interaction, respectively. $\mu$ is the chemical potential.
$d$ is the dimensionality of the system, $t^{(a)}$ the hopping
integral of the $a$-th orbital and $\alpha_{i,j}$ is the projection
operator on nearest-neighbor sites. The double occupancy operator
per orbital is defined as
$D_a(i)=n_{a,\uparrow}(i)n_{a,\downarrow}(i)$. We have also
introduced the spinorial notation
$$c_a^\dag(i)=(c_{a,\uparrow}^\dag (i), \,
c_{a,\downarrow}^\dag(i))$$ and
$c_a^\alpha(i)=\sum_j\alpha_{i,j}c_a(j)$. We will fix $U'=U$
according to symmetry considerations and use $t^{(2)}$ as energy
unit. Following the Composite Operator Method prescriptions
\cite{Mancini_04} in the pole-approximation flavor, we introduce
the projector operators $\xi_a(i)=[1-n_a(i)]c_a(i)$ and
$\eta_a(i)=n_a(i)c_a(i)$ ($a=1,\,2$) and the composite field
$\psi^\dag(i)=(\xi_1^\dag(i),\,\eta_1^\dag(i),\,\xi_2^\dag(i),\,\eta_2^\dag(i))$
as operatorial basis in order to analyze the two-orbital system
with different bandwidths ($t^{(1)} \leq t^{(2)}$). In this
approximation the Fourier transform of the retarded Green's
function $G(i,j)=\left\langle
R\left[\psi(i)\psi^\dag(j)\right]\right\rangle$ is given by
\begin{equation}
G(\mathbf{k},\omega)=\sum_m
\frac{\sigma^{(m)}(\mathbf{k})}{\omega-E_m(\mathbf{k})+\mathrm{i}\delta}
\end{equation}
The spectral functions $\sigma^{(m)}(\mathbf{k})$ and the poles
$E_m(\mathbf{k})$ can be computed \cite{Mancini_04} once the Fourier
transform of the normalization matrix
$I(\mathbf{i},\mathbf{j})=\left\langle
\left\{\psi(\mathbf{i},t),\psi^\dag(\mathbf{j},t)\right\}\right\rangle$
\begin{align}
&I=\left( \begin{array}{cc}
I^{(1)} & 0 \\
0 & I^{(2)} \\
\end{array}
\right)  \quad
n_a=\langle n_a(i)\rangle \\
&  I^{(a)}_{11}=1-\frac12 n_a \quad I^{(a)}_{12}=I^{(a)}_{21}=0
\quad I^{(a)}_{22}=\frac12 n_a
\end{align}
and of the matrix $m(\mathbf{i},\mathbf{j})=\left\langle
\left\{\mathrm{i}\frac{\partial}{\partial
t}\psi(\mathbf{i},t),\psi^\dag(\mathbf{j},t)\right\}\right\rangle$
(we here report only the non-zero entries)
\begin{align}
&m_{11}(\mathbf{k})=-\mu I^{(1)}_{11}+U'(n_2-\chi_0)\nonumber\\
&-2dt^{(1)}[\Delta^{(1)}+\alpha(\mathbf{k})(1-n_1+p^{(1)})]\\
&m_{12}(\mathbf{k})=2dt^{(1)}[\Delta^{(1)}+\alpha(\mathbf{k})(p^{(1)}-I^{(1)}_{22})]\\
&m_{22}(\mathbf{k})=(U-\mu)I^{(1)}_{22}+U'\chi_0\nonumber\\
&-2dt^{(1)}[\Delta^{(1)}+\alpha(\mathbf{k})p^{(1)}]\\
&m_{33}(\mathbf{k})=-\mu I^{(2)}_{11}+U'(n_1-\chi_0)\nonumber\\
&-2dt^{(2)}[\Delta^{(2)}+\alpha(\mathbf{k})(1-n_2+p^{(2)})]\\
&m_{34}(\mathbf{k})=2dt^{(2)}[\Delta^{(2)}+\alpha(\mathbf{k})(p^{(2)}-I^{(2)}_{22})]\\
&m_{44}(\mathbf{k})=(U-\mu)I^{(2)}_{22}+U'\chi_0\nonumber\\
&-2dt^{(2)}[\Delta^{(2)}+\alpha(\mathbf{k})p^{(2)}]
\end{align}
are known. This latter depends on six parameters. Five of them
($\mu$, $\Delta^{(a)}=\left\langle \xi_a \xi_a^{\dag\alpha}
\right\rangle-\left\langle \eta_a \eta_a^{\dag\alpha}
\right\rangle$, $p^{(a)}=\frac14 \left\langle n_{a\mu}(i)
n_{a\mu}^\alpha(i) \right\rangle - \left\langle
[c_{a\uparrow}(i)c_{a\downarrow}(i)]^\alpha c_{a\downarrow}^\dag(i)
c_{a\uparrow}^\dag(i) \right\rangle$) have been fixed by algebra
constrains \cite{Mancini_04}
\begin{align}
& n=4-2(C_{11}+C_{22}+C_{33}+C_{44})\\
& \Delta^{(a)}=C_{2a-1,2a-1}^\alpha-C_{2a,2a}^\alpha \quad (a=1,2)\\
& C_{2a-1,2a}=0 \quad (a=1,2)
\end{align}
where $n$ is the total filling, $C_{mm'}=\left\langle \psi_m
\psi_{m'}^\dag \right\rangle$ and $C_{mm'}^\alpha=\left\langle
\psi_m \psi_{m'}^{\dag\alpha} \right\rangle$ are the on-site and
the nearest-neighbor-site correlation functions, respectively,
$n_{a\mu}(i)$ is the charge ($\mu=0$) and spin ($\mu=1,\,2,\,3$)
density operator. The sixth one, $\chi_0=\frac12\left\langle
n_1(i) n_2(i) \right\rangle$, representing the interorbital charge
correlations, has been fixed through decoupling $\chi_0=\frac12
n_1 n_2$.
\begin{figure}[tbph]
\centering\includegraphics*[width=0.47\textwidth]{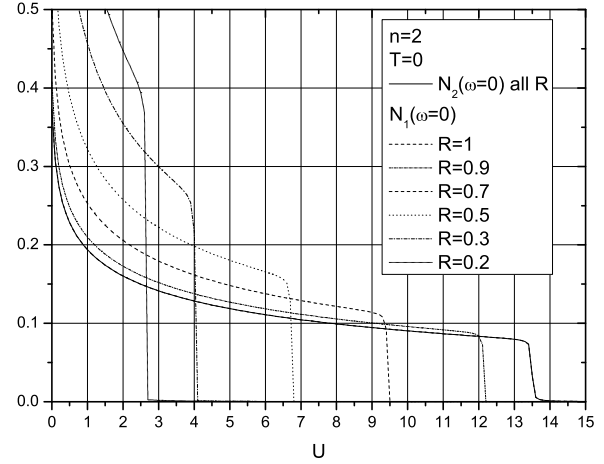}
\caption{The density of states at the chemical potential
$N_a(\omega=0)$ of both orbitals as a function of the Coulomb
potential $U$ for different values of the ratio
$R=t^{(1)}/t^{(2)}$ at $n=2$ and $T=0$.} \label{figura1}
\end{figure}
In Fig.~\ref{figura1}, we report the density of states at the
chemical potential $N(\omega=0)$ of both orbitals as a function of
the Coulomb potential $U$ for different values of the ratio
$R=t^{(1)}/t^{(2)}$ at $n=2$ and $T=0$. We can see that the
critical value of the Coulomb repulsion $U_{c2}$ at which a gap
opens in the density of states of the orbital $2$, which has full
bandwidth ($t^{(2)}=1$), remains unchanged on varying the ratio
$R$. On the contrary, $U_{c1}$, the value of the Coulomb repulsion
at which a gap opens in the density of states of the orbital $1$,
is extremely sensible to the value of the $R$ and seems to obey a
linear relationship with this latter.

In conclusion, we have shown that the Composite Operator Method is
capable to obtain an orbital selective Mott transition scenario in
the two-orbital Hubbard model already within the two-pole
approximation. We need now to improve the basis in order to get a
more realistic picture and compare our results with experiments.

\end{document}